\documentclass[prb,onecolumn,nofootinbib,citeautoscript,10pt,notitlepage]{revtex4-2}
\usepackage{dsfont}
\usepackage{float}
\usepackage{array}
\usepackage{slashed,bbold}
\usepackage{physics}
\usepackage{enumerate}

\usepackage{amsmath,amssymb,bm} 
\usepackage{graphicx}

\usepackage[tight]{subfigure} 

\usepackage{color} 
\usepackage[papersize={8.5in,11in}]{geometry}

\usepackage{color}
\definecolor{darkblue}{rgb}{0.,0.,0.4}
\definecolor{darkred}{rgb}{0.5,0.,0.}
\definecolor{BlueViolet}{RGB}{138,43,226}
\definecolor{SkyBlue}{RGB}{30,144,255}
\definecolor{DarkGreen}{RGB}{0,100,0}
\usepackage[pdftex,colorlinks=true,linkcolor=darkblue,citecolor=blue,urlcolor=darkred]{hyperref}

\renewcommand{\epsilon}{\varepsilon}

\def \be{\begin{equation}}
\def \ee{\end{equation}}
\def \nn{\nonumber \\}
\begin{document}

\title{Andreev bound states in superconductor-barrier-superconductor junctions of Rarita-Schwinger-Weyl semimetals}

\author{Ipsita Mandal}
\affiliation{Department of Physics, Shiv Nadar Institution of Eminence (SNIoE), Gautam Buddha Nagar, Uttar Pradesh 201314, India}

\begin{abstract}
We consider a superconductor-barrier-superconductor configuration built with Rarita-Schwinger-Weyl semimetal, which features four bands crossing at a single nodal point. Assuming a homogeneous s-wave pairing in each superconducting region, and the barrier region created by applying a voltage of magnitude $V_0 $ across a piece of normal state semimetal, we apply the BdG formalism to compute the discrete energy spectrum $\varepsilon $ of the subgap Andreev bound states in the short-barrier regime. In contrast with the two-band semimetals studied earlier, we find upto four pairs of localized states (rather than one pair for two-band semimetals) in the thin-barrier limit, and each value of $\varepsilon $ has a complicated dependence on the phase difference $\varphi_{12} $ via cosine and sine functions, which cannot be determined analytically. These are artifacts of multi-band nodes hosting quasiparticles of pseudospin values greater than $1/2$. Using the bound state energies, we compute the Josephson current across the junction configuration. 
\end{abstract}

\maketitle

\tableofcontents

%%%%%%%%%%%%%%%%%%%%%%%%%%%%%%%%%%%%%%%%%%%%%%%%%

\section{Introduction}

A large number of gapless topological phases has been discovered in recent years which are characterized by the Brillouin zone (BZ) harbouring pairs of points where two or more bands cross \cite{bernevig,bernevig2}. Often such systems have a nontrivial topology in the momentum space, exhibiting nonzero Chern numbers about each band-crossing point. The associated materials are called semimetals due to the existence of the gapless nodal points where the density of states goes to zero. The simplest and the most well-known three-dimensional (3d) example is the Weyl semimetal (WSM) \cite{burkov11_weyl,yang_weyl}, which exhibits an isotropic linear-in-momentum dispersion in the vicinity of two bands crossing at a point. A simple generalization of the WSM is a multiband semimetal with isotropic linear dispersions, whose low-energy effective Hamiltonian can be expressed as $\sim \mathbf{k} \cdot \boldsymbol{\mathcal{S}} $,
where $\boldsymbol{\mathcal{S}}$ represents the three-component vector consisting of the matrices for a particular value of pseudospin, with the nomenclature ``pseudospin'' being used to unambiguously differentiate it from the actual (relativistic) spin. These are the higher-pseudospin-semimetals (i.e., with pseudospin value greater than $1/2$), thus, constitute natural generalizations of the WSM Hamiltonian $\sim \mathbf{k} \cdot \boldsymbol{\sigma}$, on account of the number of bands being higher than two.\footnote{Here we have used the usual convention that $\boldsymbol{\sigma}$ represents the vector of the three Pauli matrices, implying that the WSM hosts pseudospin-1/2 quasiparticles.} Examples of multiband semimetals include the pseudospin-1 Maxwell fermions \cite{spin-1,ips3by2,ips-cd,ips-magnus} (with threefold band-crossings) and the pseudospin-3/2 Rarita-Schwinger-Weyl (RSW) semimetals~\cite{long,igor,igor2,isobe-fu,ips3by2,ips-cd,ips-magnus,ips_jns} (with fourfold band-crossings).

In the branch of high-energy physics, the Rarita-Schwinger (RS) equation describes the field equation for elementary (relativistic) particles with a spin value of 3/2. Although they are postulated to exist in models based on supergravity \cite{weinberg}, they do not appear in the standard model, and none has been detected experimentally either. On the other hand, in condensed matter systems, an analogue of these relativistic spin-3/2 fermions exists in the form of quasiparticles carrying pseudospin-3/2~\cite{long,igor,igor2,ips3by2,ips-cd,ips-magnus,ips_jns}, which of course are non-relativistic. In an effective Hamiltonian of the form shown in Eq.~\eqref{hamrsw0}, the four bands show linear-in-momentum dispersions fixed to the values $\pm 3 \,|\mathbf k|/2 $ and $\pm |\mathbf k|/2 $ [cf. Eq.~\eqref{eqdisrsw}].
%%%%%
It has been argued that the large topological charges, found in various materials like CoSi~\cite{takane}, RhSi~\cite{sanchez}, AlPt~\cite{schroter}, and PdBiSe~\cite{ding}, represent the features of an RSW semimetal.

\begin{figure}[t]
\subfigure[]{\includegraphics[width = 0.5 \textwidth]{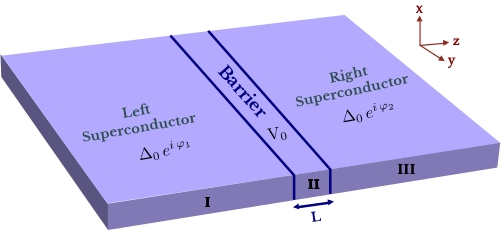}}
\hspace{3 cm}
\subfigure[]{\includegraphics[width = 0.2 \textwidth]{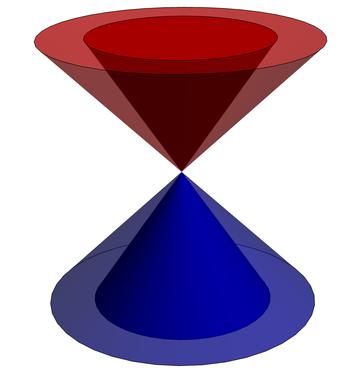}}
\caption{\label{figrsw}
Schematics of the
(a) S-B-S junction set-up;
(b) isotropic energy dispersions of an RSW semimetal, plotted against the $k_x$-$k_y$/$k_y$-$k_z$/$k_z $-$k_x $ plane, featuring conical dispersions with two distinct values of the magnitude of the slope [viz. $\mathcal{E}_{3/2}^s(\mathbf{k}) = s \, { 3\,|\mathbf k|}/ {2} $ and $\mathcal{E}_{1/2}^{s}(\mathbf{k}) = s \,{  |\mathbf k|}/ {2}$, with $s =\pm \,$].}
\end{figure}

One way to understand the consequences of the existence of nonzero pseudospin quasiparticles is to study the Josephson effect in set-ups consisting of junctions between the normal (abbreviated by ``N'') and the superconducting (abbreviated by ``S'') states of various semimetals. The superconductivity is induced via proximity-effect by placing a conventional s-wave superconductor atop the corresponding region \cite{proximity-sc}. Examples of relevant configurations include N-S~\cite{beenakker,tanaka_review,spin1,wsm_jj2,wsm_jj0,wsm_jj1}, S-N-S~\cite{titov-graphene,bolmatov_graphene_sns}, and S-B-S (where ``B'' indicates a potential barrier in the N region, which can be created by applying a gate voltage $V_0$ across the normal state region) \cite{krish-moitri,emil_jj_WSM,debabrata-krish,debabrata} junctions. These studies have considered both 2d \cite{beenakker,spin1,krish-moitri} and 3d \cite{wsm_jj2,wsm_jj0,wsm_jj1,debabrata-krish,debabrata} semimetals. Although the pseudospin-1 semimetal has three bands crossing at a point, it has a flat (i.e., nondispersive) band which does not participate in transport. Hence, we extend the earlier studies (involving pseudospin-1/2 semimetals) by considering an S-B-S set-up, as shown in Fig.~\ref{figrsw}(a), constructed out of RSW semimetals, with the superconducting region exhibiting spin-singlet s-wave pairing \cite{igor}. We focus on the short-barrier regime, such that the barrier thickness $ L $ is taken to be $ L \ll \xi $, where $\xi $ is the superconducting coherence length (i.e., the subgap excitations decay over a length scale $\sim \xi$ inside the superconductor).

Let the strength of the superconducting order parameter be given by $\Delta = \Delta_0 \, e^{i\,\varphi} $ and let $\varepsilon$ represent the energy of the emergent eigenstates. We get a set of discrete states for $ |\varepsilon| <\Delta_0 $, also known as the subgap excitations. For $|\varepsilon | >\Delta_0 $, the states form a continuum. Due to the fact that four bands cross at a single node of an RSW semimetal, it is expected to exhibit features which are distinct from the systems studied so far in the existing literature. In particular, for two-band semimetals, it has been shown \cite{titov-graphene,krish-moitri,debabrata-krish} that the energy of the Andreev bound states (ABSs) in the thin-barrier-limit is given by $ \varepsilon = \pm \, \Delta_0 \, \sqrt{1 - T_N\, \sin^2 \left( \varphi_{12} / 2\right) }$. Here, $\varphi_{12} $ is the difference of the superconducting phases on the two sides of the barrier region, and $T_N$ is the transmission coefficient in an analogous set-up with the superconducting regions replaced by the normal state of the semimetal. This result follows from the fact that the solution for $\eta \equiv \cos(2\,\varepsilon/\Delta_0 )$ is obtained from a linear equation (i.e., a first-order polynomial equation in $\eta$), whose $\eta$-independent coefficient contains a term proportional to $\cos \varphi_{12} $.
However, this result will not hold true for an RSW semimetal, where we get more than two pairs of ABSs with energies of the form $\pm |\varepsilon| $. This results from the fact that the solution for the RSW case involves a complex-valued quartic equation in the variable $\eta_R\equiv  \exp(2\,i\,\varepsilon/\Delta_0 ) $, with both $\cos \varphi_{12} $ and $\sin \varphi_{12} $ (and their products) appearing in the various coefficients of the polynomial.

We consider the propagation of quasiparticles and quasiholes in a slab of square cross-section, with a side-length $W$, where $W$ is assumed to be large enough to impose periodic boundary conditions along these transverse directions. The propagation direction of the quasiparticles/quasiholes is taken to be parallel/antiparallel to the $z$-axis. We compute the energies of the ABSs in the thin-barrier-limit, which is the limit when the strength of the potential barrier $V_0 \rightarrow  \infty $ and $L \rightarrow 0 $, with $\chi \equiv V_0\, L$ held fixed at a finite value. In the short-barrier regime, the dominant contribution to the Josephson current comes from the subgap states (i.e., the bound states populating the discrete Andreev energy levels) \cite{been_houten,titov-graphene,zagoskin}, because the contributions from the excited states in the continuum (with the magnitude of the energy $\varepsilon $ exceeding $ \Delta_0 $) are smaller by a factor of $L/\xi$ and, hence, are negligible in this limit. 

The paper is organized as follows. In Sec.~\ref{secham}, we describe the low-energy effective Hamiltonian of the RSW in its normal state, and show its eigenvalues and eigenfunctions. In Sec.~\ref{secsbs}, the S-B-S junction set-up is explained and the Bogoliubov–de Gennes (BdG) Hamiltonian is constructed. The expressions for the electron-like and hole-like wavefunctions are also elucidated there. This is followed by Sec.~\ref{secresults}, where the methodology employed to obtain the ABS spectrum is explained and some representative values are illustrated in various parameter regimes. We also numerically find the Josephson current from these bound states. Finally, we end with a summary and outlook in Sec.~\ref{secsum}.

 %%%%%%%%%%%%%%%%%%%fig 1 %%%%%%%%%%%%%%%%%%%%%%%%%%%%
%%%%%%%%%%%%%%%%%%%%%%%%

%%%%%%%%%%%%%%%%%%%%%%%%%%%%%%%%%%%%%
\section{RSW semimetal}
\label{secham}

Using symmetry analysis, it has been shown that various crystal structures, belonging to the eight space groups $207$-$214$, host fourfold topological degeneracies about the $\Gamma $, R, and/or H points \cite{bernevig}. On linearizing the $\mathbf{k} \cdot \mathbf {p}$ Hamiltonian about such a nodal point, we arrive at the effective continuum Hamiltonian, in the low-energy limit, captured by
\begin{align}
\label{hamrsw0}
\mathcal{H}_{ \text{RSW}}(\mathbf  k) 
= v \,\mathbf{k}\cdot \mathbf J\,.
\end{align}
Here, $v$ denotes the magnitude of the group velocity of the quasiparticles and $\boldsymbol{\mathcal{S}}=\mathbf J $. Henceforth, we will set $v=1 $ for the sake of simplicity.
%%%%%%%%%%%%%%%%%%%%%%%%%%%%%%
The system hosts pseudospin-3/2 RSW quasiparticles, which is reflected by the fact that
the three components of $\mathbf J$ form the spin-3/2 representation of the SO(3) group. A standard representation of $\mathbf J$ is given by
\begin{align}
\label{eqrswxJ}
J_x = \left(
\begin{array}{cccc}
 0 & \frac{\sqrt{3}}{2} & 0 & 0 \\
 \frac{\sqrt{3}}{2} & 0 & 1 & 0 \\
 0 & 1 & 0 & \frac{\sqrt{3}}{2} \\
 0 & 0 & \frac{\sqrt{3}}{2} & 0 \\
\end{array}
\right), \, \,
%%%%%%%%%%%%%%%
J_y = \left(
\begin{array}{cccc}
 0 & \frac{-\sqrt{3} \, i } {2}   & 0 & 0 \\
 \frac{ \sqrt{3}\, i }{2} & 0 & -  i  & 0 \\
 0 &  i  & 0 & \frac{ - \sqrt{3}\, i }{2}  \\
 0 & 0 & \frac{ \sqrt{3} \, i } {2} & 0 \\
\end{array}
\right)\,,\,\,
%%%%%%%%
J_z =\frac{1}{2} \left(
\begin{array}{cccc}
 3 & 0 & 0 & 0 \\
 0 & 1 & 0 & 0 \\
 0 & 0 & -1 & 0 \\
 0 & 0 & 0 & -3 \\
\end{array}
\right)\,.
\end{align}

The energy eigenvalues take the forms:
\begin{align}
\label{eqdisrsw}
\mathcal{E}_{3/2}^s(\mathbf{k}) = s \,\frac{ 3\,|\mathbf k|} {2}
\text{ and } \mathcal{E}_{1/2}^{s}(\mathbf{k}) = s \,\frac{  |\mathbf k|} {2} \,,
\text{ where } s=\pm \,,
\end{align}
demonstrating four linearly dispersing bands crossing at a point [cf. Fig.~\ref{figrsw}(b)].
Here the ``$+$" and ``$-$" signs, as usual, refer to the conduction and valence bands, respectively.
The corresponding orthonormal eigenvectors are given by
\begin{align}
\label{eqev}
& \Psi^s_{3/2} ( \mathbf k) = \frac{1} {\mathcal{N}^s_{3/2} }
\left [
\frac{ s\,k \left( k_x^2+k_y^2+4\, k_z^2\right)+
k_z \left(3\, k_x^2+3 k_y^2+4 \,k_z^2\right)} { \left (k_x+   i  \, k_y \right )^3}
\quad
\frac{\sqrt{3} \left [ 2\, k_z \left (s\, k+k_z \right )+k_x^2+k_y^2\right ]}
{\left (k_x+  i  \,k_y\right )^2}
\quad
\frac{\sqrt{3} \left (s\, k+k_z \right ) }
{k_x+  i \,k_y}
\quad 1   \right]^T
\nn & \hspace{0 cm} \left(\text{for energy } \mathcal{E}_{3/2}^{s} \right)
\nn
%%%%%%%%%%%%%%%%%%%
& \text{and }
\Psi^s_{1/2}  ( \mathbf k) = \frac{1} { \mathcal{N}^s_{1/2} }
\left[ \frac{ - \left( s\, k+k_z  \right )  \left (k_x-  i \, k_y \right )}
{(k_x+  i \,k_y)^2}
\quad
 \frac{2 \,k_z \left ( s\, k+k_z \right ) -k_x^2 - k_y^2}
{\sqrt{3} \left (k_x+  i \, k_y \right )^2}
\quad \frac{ s\, k+3 \, k_z}{\sqrt{3} 
\left (k_x+  i \, k_y \right )}
\quad  1  \right ] ^T
\left(\text{for energy } \mathcal{E}_{1/2}^{s} \right),
\end{align}
where $ k=\sqrt{k_x^2+k_y^2+k_z^2}$, and $ {1} / {\mathcal{N}^s_{3/2} }$ and $ {1} / {\mathcal{N}^s_{1/2}}$ denote the corresponding normalization factors. If the Fermi energy cuts the bands at energy $E$, then for propagation along the $z$-direction, the corresponding plane waves will have the factors $e^{i\, \text{sgn}(E) \,k_z^{(3/2)}\, z}$ and $e^{i\, \text{sgn}(E) \,k_z^{(1/2)}\, z}$, such that $ k_z^{(3/2)}  = \sqrt{ \left( \frac{E}{3/2}  \right)^2-k_x^2 -k_y^2 }$ and $ k_z^{(1/2)}  = \sqrt{ \left( \frac{E}{3/2}  \right)^2-k_x^2 -k_y^2 } \,$.

%%%%%%%%%%%%%%%%
\section{S-B-S junction}
\label{secsbs}

The main influence of the presence of pseudospin-$3/2 $ quasiparticles is the existence of four bands, giving rise to four independent wavefunctions for the normal-phase Hamiltonian. The first task is to determine what kind of Cooper pairing can arise in such a four-band system, with each band featuring an isotropic linear-in-momentum dispersion. This issue has been addressed in Ref.~\cite{igor}, which requires some symmetry analysis. The author demonstrates that the s-wave
superconducting state, represented by the order parameter
\begin{align}
\Delta_{sc}  = \langle \psi^\dagger \, \Gamma \,\psi^* \rangle,
\end{align} 
where $\psi$ is the four-component quasiparticle spinor,
opens a Majorana mass gap for the fermions and is the leading superconducting
instability. The explicit form of $\Gamma $ is shown in Eq.~\eqref{eqscorder} presented below.

In order to get the S-B-S configuration [cf. Fig.~\ref{figrsw}(a)], we model the superconducting pair potential as \cite{igor}
\begin{align}
\label{eqscorder}
\Delta (z) =\begin{cases} 
\Delta_0\,e^{i\,\varphi_1 }\,\Gamma
 &\text{ for }  z\leq 0   \\
0 &\text{ for } 0 < z< L \\
\Delta_0\,e^{i\,\varphi_2 } \,\Gamma &\text{ for } z \geq L
\end{cases}, \quad
%%%%%%%%%%%%%%%
\Gamma= i \left( \frac{J_y \, J_z +J_z \, J_y } {\sqrt 3} \right)
\left( \frac{J_x \, J_y + J_y \, J_x } {\sqrt 3} \right),
\end{align}
representing Cooper pairing in the s-wave channel. Due to the presence of the barrier region, we need to consider the potential energy
\begin{align}
V(z)
=\begin{cases} 
0
 &\text{ for }  z\leq 0 \text{ and } z\geq L  \\
 V_0 &\text{ for } 0 < z< L 
\end{cases}.
\end{align} 
%%%%%%%%%%%%%%%%%%%%%%%%
The resulting BdG Hamiltonian is given by
% from PRB 98, 224509 (2018)
\begin{align}
& H = \frac{1} {2} \sum_{\mathbf k} \Psi^\dagger_{\mathbf k} \,H_{\text{BdG}} (\mathbf k)
\Psi_{\mathbf k} , \quad
\Psi_{\mathbf k} = \begin{bmatrix}
c_1 (\mathbf k) & c_2 (\mathbf k) & c_3 (\mathbf k) & c_4 (\mathbf k)  & 
c_1^\dagger (-\mathbf k) & c_2 ^\dagger(-\mathbf k) & c_3^\dagger (-\mathbf k) & c_4^\dagger (-\mathbf k)
\end{bmatrix}^T, \nn
%%%%%%%%%%%%%%%%%%%%%%%%%%%
& H_{\text{BdG}} (\mathbf k) =
\begin{bmatrix}
\mathcal{H}_{ \text{RSW}}(\mathbf  k) -E_F + V(z)  & \Delta(z)  \\ 
 \Delta^\dagger(z) &  E_F- V(z) -\mathcal{H}_{ \text{RSW}}^T(-\mathbf  k)  \\  
\end{bmatrix},
%%%%%%%%%%%%%%%%%%%%%%%
\label{eq_bdg}
\end{align}
where the subscripts $\lbrace 1, 2, 3, 4\rbrace $ on the fermionic operators represent the four distinct band indices.
Here we demarcate the left superconducting region as ``region I'', the middle barrier region as ``region II'', and the right superconducting region as ``region III''. The electron-like and the hole-like BdG quasiparticles are obtained from the eigenvalue equation
\begin{align}
H_{\text{BdG}} ( \mathbf k\rightarrow -i \boldsymbol{\nabla}_{\mathbf r}) \,\psi_{\mathbf k} (\mathbf r) 
= \varepsilon \, \psi_{\mathbf k} (\mathbf r) \,,
\end{align}
where $\mathbf r=(x, \, y,\, z)$ is the position vector. If $ \psi_N (\mathbf k)$ is an eigenfunction of $\mathcal{H}_{ \text{RSW}}(\mathbf  k)$ (with the superconducting phase factor of $\varphi $), then the electron-like and hole-like eigenfunctions of $ H_{\text{BdG}} ( \mathbf k) $ are given by the expressions~\cite{timm}
\begin{align}
\label{eqelechole}
\psi_e (\mathbf k) = \begin{bmatrix}
\psi_N (\mathbf k) & \frac{ \left( \epsilon -\Omega \right) \, e^{-i\,\varphi }} {\Delta_0}\,
\Gamma \cdot \psi_N (\mathbf k)
\end{bmatrix} \text{ and }
%%%%%%%%%%%%
\psi_h (\mathbf k) = \begin{bmatrix}
\psi_N (\mathbf k) & \frac{ \left( \epsilon + \Omega \right) \, e^{-i\,\varphi }} {\Delta_0}\,
\Gamma \cdot \psi_N (\mathbf k)
\end{bmatrix} ,
\end{align}
respectively, where 
\begin{align}
\Omega = i \,\sqrt{ \Delta_0^2 - \epsilon^2 } \,.
\end{align}

Let us define
\begin{align}
\beta =\arccos(\varepsilon /\Delta_0)\,,
\end{align}
which will be useful in the expressions that follow.
%%%%%%%%%%%%%%%%%%%%%%%%%%%%%%%
% See Asano lectures, pg-54
Using Eqs.~\eqref{eqev} and \eqref{eqelechole}, let us now elucidate the form of the eigenfunction
$$ \Psi (\mathbf r, k_\perp) = 
\psi_{I} (\mathbf r, k_\perp) \,\Theta(-z)
 + \psi_{II} (\mathbf r , k_\perp) \,\Theta(z)\, \Theta(L-z) 
 +  \psi_{III} (\mathbf r , k_\perp) \,\Theta(z-L) \,,$$
expressed in a piecewise manner for the three regions, where we set the the Fermi energy at $E_F$ for the corresponding normal states (i.e., for $\Delta_0 = 0 $) in the regions I and III. We assume the energy-scale hierarchies $V_0 \gg E_F  \gg \Delta_0 $ and $(V_0-E_F) \gg E_F $.\footnote{The condition $\Delta_0 \ll E_F$ ensures that the mean-field approximation, applicable for using the BdG formalism, is valid. The second condition $ (V_0-E_F) \gg E_F $ arises because we are focussing on the short-barrier regime.} Since the propagation direction is along the $z$-axis, the translation symmetry is broken in that direction, whereas the transverse momentum components $k_x$ and $k_y $ are conserved across the S-B and B-S junctions.
We denote the magnitude of the transverse component as $k_\perp =\sqrt{k_x^2 + k_y^2 } $, and the azimuthal angle $\phi =\arctan(k_y /k_x )$.

%%%%%%%%%%%%%%%%%%%%%%%%%%%%%%%%%%%%%%%%%%%%%%%%%%%%%%
\begin{enumerate}

\item In the right superconductor region, the wavefunction localizing at the
interface is described by a linear combination of the following form (see chapter 5 of Ref.~\cite{asano}):
% rt moving holes: we use nomenclature of Phys. Rev. B 97, 041116 (2018)
\begin{align}
\psi_{III} ( \mathbf r, k_\perp ) = a_{32r}\,\psi_{3/2}^{er}  ( \mathbf r, \theta _{32}^r )
 + a_{12r}\,\psi_{1/2}^{er}  ( \mathbf r,\theta_{12}^r )
+ b_{32r}\,\psi_{3/2}^{hr}  ( \mathbf r, \tilde \theta_{32}^r ) 
+ b_{12r}\,\psi_{1/2}^{hr}   ( \mathbf r, \tilde \theta_{12}^r )\,,
\end{align}
where

%%%%%%%%%%%%%%%%%%%%%%%%%%%%%
\begin{align}
 \psi_{3/2}^{er}   ( \mathbf r, \theta _{32}^r ) &= \frac{ 
e^{ i \left \lbrace  k_x\, x \, + \, k_y\, y \, 
+ \, k_{z}^{(3/2),el }\, (z-L) \right \rbrace  } \,
e^{- i \,\phi \, {\mathbb{1}_{2\times 2}} \otimes J_z}
}
{\sqrt 2}
\nn &  \qquad  \times
%%%%%%%%%%%%%%%%%%%%%%%
\Big [\begin{matrix}
\frac{ e^{i \beta } \cos ^3 ( \frac{\theta _{32}^r}{2} )}
{\sqrt{3}}
& \frac{ e^{i \beta } 
\sin   \theta _{32}^r  \cos  ( \frac{\theta _{32}^r}{2} )
}
{2}   
 & \frac{ e^{i \beta }  \sin  ( \frac{\theta _{32}^r}{2} ) \sin   \theta _{32}^r }
 {2}   
& \frac{ e^{i \beta } \sin ^3 ( \frac{\theta _{32}^r}{2} )} {\sqrt{3}}
\end{matrix}
%%%%%%%%%%%%%%%%%%%%%%%%%%%%%%%%
\nn  & \hspace{ 3.5 cm} \begin{matrix}
\frac{-i \, e^{-i\,   \varphi_2} \sin ^3 ( \frac{\theta _{32}^r}{2} )} {\sqrt{3}}  
&
\frac{i  \, e^{-i\,   \varphi_2}
\sin  ( \frac{\theta _{32}^r}{2} ) \sin   \theta _{32}^r 
}
{2}
&
\frac{ - i  \, e^{-i\,   \varphi_2}
\sin  \theta _{32}^r  \cos  ( \frac{\theta _{32}^r}{2} )
}
{2}  
& \frac{i \, e^{-i\,   \varphi_2} \cos ^3 ( \frac{\theta _{32}^r}{2} )} {\sqrt{3}}
\end{matrix} \Big ]^T ,\nn
%%%%%%%%%%%%%%%%%%%%%%%
\sin  \theta _{32}^r  \simeq  \frac{3 \,k_\perp /2} {E_F} & \,, \quad
k_{z}^{(3/2),er } \simeq  k^{(3/2)}_{\rm{mod} }
+ i \,\kappa_1 \,,\quad
k^{(3/2)}_{\rm{mod} } \simeq 
\sqrt{ \left( \frac{E_F} {3/2} \right) ^2  - k_\perp^2 } \,,
\quad 
\kappa_1 = \frac{E_F \,\Delta_0 \sin \beta } 
{ (3/2)^2 \,k^{(3/2)}_{\rm{mod} } } \,,
\quad
\tan \theta _{32}^r  \simeq \frac{k_\perp} { k^{(3/2)}_{\rm{mod} }}\,,
\end{align}
%%%%%%%%%%%%%%%%%%%%%%%%%%%%%

%%%%% electron-like 1/2 %%%%%%%%%%%%%%%%%%555
\begin{align}
  \psi_{1/2}^{er}  ( \mathbf r, \theta _{12}^r )
 & = \frac{ 
e^{ i \left \lbrace k_x\, x \, + \, k_y\, y \, + \, k_{z}^{(1/2),el }\, 
(z-L) \right \rbrace  
} \,
e^{- i \,\phi \, {\mathbb{1}_{2\times 2}} \otimes J_z}
}
{\sqrt 2}
\nn &  \, \times
%%%%%%%%%%%%%%%%%
\begin{matrix}
\Big [
\frac{  -\sqrt{3}  \, e^{i \beta } 
\sin \theta _{12}^r  \cos (\frac{\theta _{12}^r}{2} )
}
{2} 
%%%%%%%%%%
& \frac{  
e^{i \beta } \cos  (\frac{\theta _{12}^r}{2} ) 
\left(3 \cos \theta _{12}^r-1\right)
}
{2} 
&
\frac{e^{i \beta } \sin  (\frac{\theta _{12}^r}{2} ) 
\left(3 \cos  \theta _{12}^r +1 \right)}
{2} 
%%%%%%%
&
\frac{ \sqrt{3}\, e^{i \beta } \sin (\frac{\theta _{12}^r}{2} ) 
\sin  \theta _{12}^r
}
{2} 
\end{matrix} \nn &  \hspace{ 0.5 cm} 
%%%%%%%%%%%%%%%%%%%%%%%%%%%%%%%%%%%%
\begin{matrix}
\frac{ - i \,\sqrt{3} \,e^{-i \varphi_2}
\sin (\frac{\theta _{12}^r}{2} )
 \sin  \theta _{12}^r
}
{2}  
 & \frac{ i \,e^{-i \varphi_2} 
 \sin  (\frac{\theta _{12}^r}{2} ) 
 \left(3 \cos \theta _{12}^r +1\right)
 }
 {2} 
%%%%%%%%% 
 &
 \frac{i \, e^{-i \varphi_2} \cos (\frac{\theta _{12}^r}{2} )
  \left(1-3 \cos \theta _{12}^r \right)
 }
 {2}  
&
\frac{- i\, \sqrt{3}\, e^{-i \varphi_2} 
\sin \theta _{12}^r 
\cos (\frac{\theta _{12}^r}{2} )
}
{2} 
\end{matrix} \Big ]^T , \nn
%%%%%%%%%%%%%%%%%%%%%%
\sin  \theta _{12}^r  \simeq  \frac{ k_\perp /2} {E_F} 
& \,, \quad
k_{z}^{(1/2),er }\simeq  k^{(1/2)}_{\rm{mod} }
+ i \,\kappa_2 \,,\quad
k^{(1/2)}_{\rm{mod} } \simeq 
\sqrt{ \left( \frac{E_F} {1/2} \right) ^2  - k_\perp^2 } \,,
\quad 
\kappa_2 = \frac{E_F \,\Delta_0 \sin \beta } 
{ (1/2)^2 \,k^{(1/2)}_{\rm{mod} } } \,,
\quad
\tan \theta _{12}^r  \simeq \frac{k_\perp} { k^{(1/2)}_{\rm{mod} }}\,,
\end{align}
%%%%%%%%%%%%%%%%%%%%%%%%%%5

%%%%%%% 3/2 hole %%%%%%%%%%%%%%%%%55
\begin{align}
 \psi_{3/2}^{hr}   ( \mathbf r, \tilde \theta _{32}^r ) & = \frac{ 
e^{ i \left \lbrace k_x\, x \, + \, k_y\, y \, + \, k_{z}^{(3/2), hl }\, (z-L) \right \rbrace 
 } \,
e^{- i \,\phi \, {\mathbb{1}_{2\times 2}} \otimes J_z}
}
{\sqrt 2}
\nn &  \qquad  \times
%%%%%%%%%%%%%%%%%%%%%%%
\Big [\begin{matrix}
\frac{ e^{i \varphi_2 } \cos ^3 ( \frac{ \tilde \theta _{32}^r}{2} )}
{\sqrt{3}}
& \frac{  e^{i \varphi_2 }
\sin   \theta _{32}^r  \cos  ( \frac{  \tilde  \theta _{32}^r}{2} )
}
{2}   
 & \frac{ e^{i \varphi_2 }  \sin  ( \frac{  \tilde  \theta _{32}^r}{2} ) \sin   \theta _{32}^r }
 {2}   
& \frac{ e^{i \varphi_2 }
\sin ^3 ( \frac{  \tilde  \theta _{32}^r}{2} )} {\sqrt{3}}
\end{matrix}
%%%%%%%%%%%%%%%%%%%%%%%%%%%%%%%%
\nn  & \hspace{ 3.5 cm} \begin{matrix}
\frac{-i \,e^{i \beta }   \sin ^3 ( \frac{  \tilde  \theta _{32}^r}{2} )} {\sqrt{3}}  
&
\frac{i  \, e^{i \beta } 
\sin  ( \frac{  \tilde  \theta _{32}^r}{2} ) \sin   \tilde  \theta _{32}^r 
}
{2}
&
\frac{ - i  \, e^{i \beta } 
\sin   \tilde  \theta _{32}^r  \cos  ( \frac{  \tilde  \theta _{32}^r}{2} )
}
{2}  
& \frac{i \, e^{i \beta }  \cos ^3 ( \frac{  \tilde  \theta _{32}^r}{2} )} {\sqrt{3}}
\end{matrix} \Big ]^T ,\nn
%%%%%%%%%%%%%%%%%%%%%%%%%%%%%%%%%
\sin  \tilde \theta _{32}^r & \simeq  \frac{3 \,k_\perp /2} {E_F} \,, \quad
k_{z}^{(3/2),hr } \simeq  -k^{(3/2)}_{\rm{mod} }
+ i \,\kappa_1 \,,
\quad
\tan \tilde \theta _{32}^r  \simeq \frac{k_\perp} { -k^{(3/2)}_{\rm{mod} }} \,,
\end{align}
%%%%%%%%%%%%%%%%%%%%%%%%%%%%%

%%%%%%%%%%%%%%%%%%%%% 1/2 hole %%%%%%%%%%%%%%%%555
\begin{align}
  \psi_{1/2}^{hr}  ( \mathbf r , \tilde \theta _{12}^r)
 & = \frac{ 
e^{ i \left \lbrace k_x\, x \, + \, k_y\, y \, + \, k_{z}^{(1/2),hl }\,
(z-L) \right \rbrace  } \,
e^{- i \,\phi \, {\mathbb{1}_{2\times 2}} \otimes J_z}
}
{\sqrt 2}
\nn &  \, \times
%%%%%%%%%%%%%%%%%
\begin{matrix}
\Big [
\frac{  -\sqrt{3}  \, e^{i \varphi_2}
\sin  \tilde \theta _{12}^r  \cos (\frac{ \tilde \theta _{12}^r}{2} )
}
{2} 
%%%%%%%%%%
& \frac{  
e^{i \varphi_2}  \cos  (\frac{ \tilde \theta _{12}^r}{2} ) 
\left(3 \cos  \tilde \theta _{12}^r-1\right)
}
{2} 
&
\frac{ e^{i \varphi_2} \sin  (\frac{ \tilde \theta _{12}^r}{2} ) 
\left(3 \cos  \tilde \theta _{12}^r +1 \right)}
{2} 
%%%%%%%
&
\frac{ \sqrt{3}\, e^{i \varphi_2} \sin (\frac{ \tilde \theta _{12}^r}{2} ) 
\sin   \tilde \theta _{12}^r
}
{2} 
\end{matrix} \nn &  \hspace{ 0.5 cm} 
%%%%%%%%%%%%%%%%%%%%%%%%%%%%%%%%%%%%
\begin{matrix}
\frac{ - i \,\sqrt{3} \, e^{i \beta}
\sin (\frac{ \tilde \theta _{12}^r}{2} )
 \sin   \tilde \theta _{12}^r
}
{2}  
 & \frac{ i \, e^{i \beta} 
 \sin  (\frac{ \tilde \theta _{12}^r}{2} ) 
 \left(3 \cos  \tilde \theta _{12}^r +1\right)
 }
 {2} 
%%%%%%%%% 
 &
 \frac{i \, e^{i \beta} \cos (\frac{ \tilde \theta _{12}^r}{2} )
  \left(1-3 \cos  \tilde \theta _{12}^r \right)
 }
 {2}  
&
\frac{- i\, \sqrt{3}\, e^{i \beta}
\sin  \tilde \theta _{12}^r 
\cos (\frac{ \tilde \theta _{12}^r}{2} )
}
{2} 
\end{matrix} \Big ]^T , \nn
%%%%%%%%%%%%%%%%%%%%%%
\sin  \tilde \theta_{12}^r & \simeq  \frac{ k_\perp /2} {E_F} \,, \quad
k_{z}^{(1/2),hr } \simeq  -k^{(1/2)}_{\rm{mod} } + i \,\kappa_2\,,
\quad
\tan \tilde \theta _{12}^r  \simeq \frac{k_\perp} { -k^{(1/2)}_{\rm{mod} }}\,.
\end{align}

The expressions above represent right-moving electron-like and hole-like wavefunctions (using the nomenclature from Sec. S2 of Ref.~\cite{wsm_jj1}).
The expressions for the various angles and the $z$-components of the momenta, shown above, are valid in the limit $\Delta_0 \ll E_F $, which we have assumed to hold true.
Clearly, in this regime, we find that $ \tilde \theta_{32}^r \simeq \pi - \theta_{32}^r$ and  $ \tilde \theta_{12}^r \simeq \pi - \theta_{12}^r$. 
% cf. https://en.wikipedia.org/wiki/Delta_potential
The fact that the ``right-moving'' wavefunctions are the admissible ones in this region
follows because, when we solve for bound-state-problems in quantum mechanics
(for example, a Schr\"{o}dinger particle tunneling through a Dirac delta
potential barrier), we get both exponentially decaying and increasing wavefunctions
--- but to get physically admissible solutions, we retain only the decaying
ones.

%%%%%%%%%%%%%%%%%%%%%%%%%%%%Normal%%%%%%%%%%%%%%%%%%%%%%%%%%%%%%%
\item

In the normal state region, we will have a linear combination of the following form:
\begin{align}
\psi_{II}   ( \mathbf r, k_\perp ) & = a_{32}\,\psi_{3/2}^{e+}   ( \mathbf r, \theta_{32n} )  
+ b_{32}\,\psi_{3/2}^{e-}  ( \mathbf r , \theta_{32n} )
+ a_{12}\,\psi_{1/2}^{e+}  ( \mathbf r , \theta_{12n} )
+ b_{12}\,\psi_{1/2}^{e-}  ( \mathbf r, \theta_{12n} )
 \nn &\quad 
+ c_{32}\,\psi_{3/2}^{h+}  ( \mathbf r , \tilde \theta_{32n} ) 
+ d_{32}\,\psi_{3/2}^{h-}  ( \mathbf r, \tilde \theta_{32n} )
+ c_{12}\,\psi_{1/2}^{h+}  ( \mathbf r, \tilde \theta_{12n} )
 + d_{12}\,\psi_{1/2}^{h-} ( \mathbf r, \tilde \theta_{12n}  ) \,,
\end{align}
where

%%%%%%%%%%%%%%%%%%
%%%%%%%%%%%%%%%%%%%%%%%%%% electron %%%%%%%%%%%%5
\begin{align}
\psi_{3/2}^{e+}  ( \mathbf r , \theta_{32n}) &= 
e^{i \left(  k_x\, x \, + \, k_y\, y \, + \, k_{z}^{(3/2),e}\, z \right) } \, 
f_1(\theta_{32n})    ,\nn
%%%%%%%%%%%%%%%
f_1(\theta_{32n}) & = e^{- i \,\phi \,{\mathbb{1}_{2\times 2}} \otimes J_z}
\begin{bmatrix}
-\sin ^3 (\frac{ \theta_{32n} }{2} ) &
\frac{\sqrt{3}  \, \sin (\frac{ \theta_{32n} }{2}) \sin \theta_{32n}}
{2}  
& -\frac{\sqrt{3} \, \sin \theta_{32n} \cos  (\frac{ \theta_{32n} }{2} )
}
{2}  
& \cos ^3 (\frac{ \theta_{32n} }{2} )
& 0 & 0 & 0 & 0
\end{bmatrix}^T , \nn
%%%%%%%%%%%%%%%%
\psi_{3/2}^{e-}    ( \mathbf r , \theta_{32n})
&=  e^{i \left(  k_x\, x \, + \, k_y\, y \, -\, k_{z}^{(3/2),e}\, z \right) } \, 
f_1  ( \pi-\theta_{32n})\,,  
%%%%%%%%%%%%%%%%%%%%%%
\nn k_{z}^{(3/2),e} & 
%=-\frac{ ( V_0 -E_F -\epsilon )  } {3/2}  \cos \theta
= -\sqrt{ \left( \frac{  V_0 -E_F -\epsilon  }
{3/2} \right )^2  - k_\perp^2 }, 
\quad \cos \theta_{32n} =\frac{ k_{z}^{(3/2),e}} 
{2 \,(\epsilon+E_F -V_0) /3} \,,\quad
\sin  \theta_{32n} =
 \frac{ k_\perp} {2 \,(\epsilon+E_F -V_0) /3}\,,
\end{align}

%%%%%%% electron 1/2 %%%%%%%%%%%%%%%%%%%%%%%55

\begin{align}
& \psi_{1/2}^{e+}  ( \mathbf r , \theta_{12n}) = 
 e^{ i \left ( k_x\, x \, + \, k_y\, y \, +\, k_z^{(1/2),e}\, z \right) } \,
 f_2( \theta_{12n}) \,,\nn
%%%
& f_2( \theta_{12n})  = 
 e^{- i \,\phi \, {\mathbb{1}_{2\times 2}} \otimes J_z}
\begin{bmatrix}
\frac{
 \sqrt{3} \sin  (\frac{ \theta_{12n} }{2} ) \sin  \theta_{12n}
 } {2} &
\frac{ - \sin  (\frac{ \theta_{12n} }{2} ) (3 \cos  \theta_{12n}+1)
} {2} &
\frac{ \cos  (\frac{ \theta_{12n} }{2} ) (3 \cos  \theta_{12n}-1)
}  {2} &
\frac{ \sqrt{3} \sin  \theta_{12n} \cos  (\frac{ \theta_{12n} }{2} )
} {2} &
& 0 & 0 & 0 & 0
\end{bmatrix} ^T    ,\nn
%%%%%%%%%%%%%%%%
& \psi_{1/2}^{e-}   ( \mathbf r , \theta_{12n}) = 
 e^{ i \left (k_x\, x \, + \, k_y\, y \, -\, k_{z}^{(1/2),e}\, z\right)} \,
f_2(\pi-\theta_{12n})\,  ,
%%%%%%%%%%%%%%%%%%%%%%
\nn & k_{z}^{(1/2),e} 
= -\sqrt{ \left( \frac{  V_0 -E_F -\epsilon  }
{1/2} \right )^2  - k_\perp^2 }, 
\quad \cos \theta_{32n} =\frac{ k_{z}^{(1/2),e}} 
{2 \,(\epsilon+E_F -V_0) } \,,\quad
\sin  \theta_{32n} =
 \frac{ k_\perp} {2 \,(\epsilon+E_F -V_0) }\,,
\end{align}
%%%%%%%%%%%%%%%%%%%%%%%%%%

%%%%%%%%%%%% hole  3/2 %%%%%%%%%%%%%%%%%%%%%%%%%
\begin{align}
& \psi_{3/2}^{h+}   ( \mathbf r, \tilde  \theta_{32n}  )   = 
e^{i \left(  k_x\, x \, + \, k_y\, y \, +\, k_{z}^{(3/2),h }\, z \right) } \,
f_3 (\tilde \theta_{32n}) \,,\nn
&
f_3 (\tilde \theta_{32n}) =
e^{- i \,\phi \,{\mathbb{1}_{2\times 2}} \otimes J_z}
%%%%%%%%%%%%%%%%%%%
\begin{bmatrix}
 0 & 0 & 0 & 0 &
\cos ^3 (\frac{ { \tilde  \theta_{32n}} }{2} )
& \frac{\sqrt{3}  \sin  { \tilde  \theta_{32n}} 
\cos (\frac{ { \tilde  \theta_{32n}} }{2} ) 
}  
{2} 
& \frac{\sqrt{3} \sin  (\frac{ { \tilde  \theta_{32n}} }{2} ) \sin  { \tilde  \theta_{32n}}}
{2}  
& \sin ^3 (\frac{ { \tilde  \theta_{32n}} }{2} )
\end{bmatrix}^T  ,\nn
%%%%%%%%%%%%%%%%
& \psi_{3/2}^{h-}   ( \mathbf r, \tilde  \theta_{32n} )   = 
e^{i \left(  k_x\, x \, + \, k_y\, y \, -\, k_{z}^{(3/2),h }\, z \right) } \, 
f_3 ( \pi - \tilde \theta_{32n})\,,
%%%%%%%%%%%%%%%%%%%%%%
\nn & k_{z}^{(3/2),h}  
= \sqrt{ \left( \frac{  V_0 -E_F + \epsilon }
{3/2} \right )^2  - k_\perp^2 }, 
\quad  \cos \tilde \theta_{32n} =\frac{ k_{z}^{(3/2),h}} 
{2 \,(\epsilon-E_F +V_0) /3} \,,\quad
\sin  \tilde \theta_{32n} =
 \frac{ k_\perp} {2 \,(\epsilon - E_F + V_0) /3}\,,
\end{align}
%%%%%%%%%%%%%%%%%%%%%%%%%%%%%%%%%

%%%%%%%%%%%%%%%%figure %%%%%%%%%%%%%%
\begin{figure}[t]
\centering
\subfigure[]{\includegraphics[width=0.4 \textwidth]{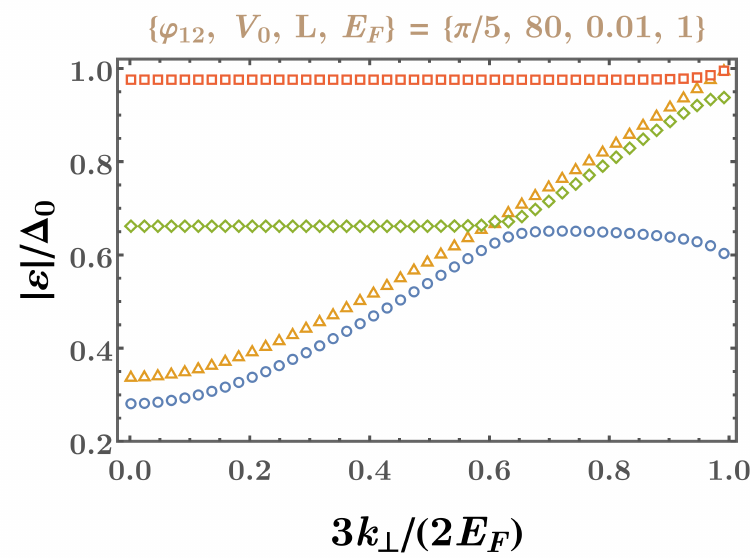}}\qquad
\subfigure[]{\includegraphics[width=0.4 \textwidth]{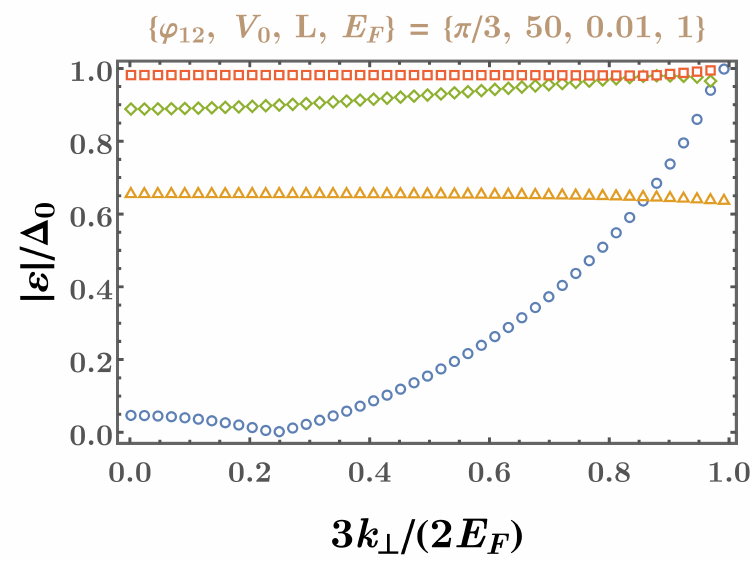}}
\subfigure[]{\includegraphics[width=0.4 \textwidth]{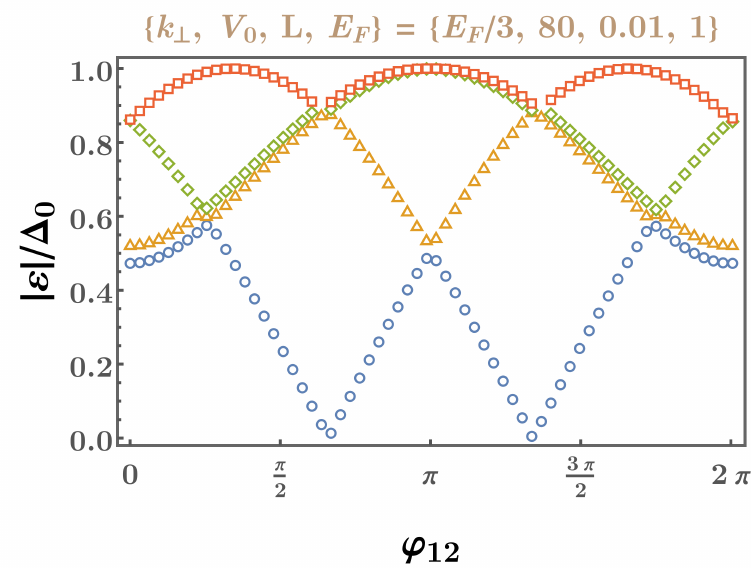}}\qquad
\subfigure[]{\includegraphics[width=0.4 \textwidth]{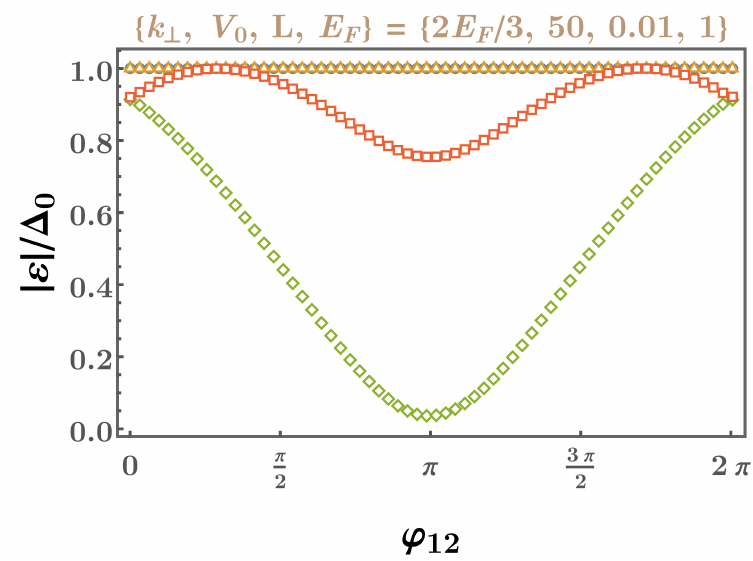}}
\caption{
Behaviour of $ |\varepsilon| $ against $k_\perp $ [subfigures (a) and (b)] and $\varphi_{12}$ [subfigures (c) and (d)], for some representative values of the remaining parameters (shown in the plotlabels).
\label{fige}}
\end{figure}
%%%%%%%%%%%%%%%%%%%%%%%%%%%%%%%%%%%%%%%%%%%%%%%%%%%

%%%% hole 1/2 %%%%%%%%%%%%%%%%%%55
\begin{align}
& \psi_{1/2}^{h+}   ( \mathbf r , \tilde \theta_{12n})   = 
e^{ i \left (k_x\, x \, + \, k_y\, y \, +\,  k_{z}^{(1/2),h}\, z \right)} \,
f_4 (\tilde \theta_{12n})\,,\nn
& f_4 (\tilde \theta_{12n}) 
= e^{- i \,\phi \, {\mathbb{1}_{2\times 2}} \otimes J_z}
%%%%%%%%%%%%%%%%%%%%%%%%%%%%%%%%%
\begin{bmatrix}
0 & 0 & 0 & 0 &
 \frac{- \sqrt{3}   \sin  { \tilde  \theta_{12n}} 
 \cos  (\frac{ { \tilde  \theta_{12n}} }{2} )} 
 {2} &
\frac{ \cos (\frac{ { \tilde  \theta_{12n}} }{2} ) (3 \cos  { \tilde  \theta_{12n}}-1)
} {2} &
 \frac{ \sin  (\frac{ { \tilde  \theta_{12n}} }{2} ) (3 \cos  { \tilde  \theta_{12n}}+1)
 }  {2} &
\frac{ \sqrt{3} \sin  (\frac{ { \tilde  \theta_{12n}} }{2} ) \sin  { \tilde  \theta_{12n}} 
} {2}
\end{bmatrix} ^T   ,\nn
%%%%%%%%%%%%%%%%
& \psi_{1/2}^{h-}  ( \mathbf r, \tilde \theta_{12n} )  = 
 e^{ i \left ( k_x\, x \, + \, k_y\, y \, -\,  k_{z}^{(1/2),h}\, z \right) } \,
 f_4 (\pi -  \tilde \theta_{12n})\,,
%%%%%%%%%%%%%%%%%%
\nn & 
k_{z}^{(1/2),h} 
= \sqrt{ \left( \frac{  V_0 -E_F + \epsilon }
{1/2} \right )^2  - k_\perp^2 }, 
\quad  \cos \tilde \theta_{12n} =\frac{ k_{z}^{(1/2),h}} 
{2 \,(\epsilon-E_F +V_0) } \,,\quad
\sin  \tilde \theta_{12n} =
 \frac{ k_\perp} {2 \,(\epsilon - E_F + V_0) }\,.
\end{align}
%%%%%%%%%%%%%%%

\item In the left superconductor region, we will have a linear combination of the following form:
\begin{align}
\psi_{I}  ( \mathbf r, k_\perp ) = a_{32l}\,\psi_{3/2}^{el}  ( \mathbf r, \theta_{32}^r )
 + a_{12l}\,\psi_{1/2}^{el}  ( \mathbf r, \theta_{12}^r )
+ b_{32l}\,\psi_{3/2}^{hl}  ( \mathbf r, \tilde \theta_{32}^r ) 
+ b_{12l}\,\psi_{1/2}^{hl}   ( \mathbf r , \tilde \theta_{12}^r )\,,
\end{align}
where
%%%%%%%%%%%%%%%%%%%%%%%%%%%%%
\begin{align}
 & \left \lbrace \psi_{3/2}^{el} ( \mathbf r, \theta_{32}^r ), \, \psi_{1/2}^{el} ( \mathbf r, \theta_{32}^r), \, 
\psi_{3/2}^{hl} ( \mathbf r, \tilde \theta_{32}^r ), 
\, \psi_{1/2}^{hl} ( \mathbf r,\tilde \theta_{32}^r ) \right \rbrace 
%%%%%%%%%%%%%%%%%%%
\nn &
= \left \lbrace \psi_{3/2}^{er} ( \mathbf r, \pi- \theta_{32}^r ), 
\, \psi_{1/2}^{er} ( \mathbf r, \pi- \theta_{12}^r ), \,
 \psi_{3/2}^{hr} ( \mathbf r, \pi-\tilde  \theta_{32}^r ) , \,
 \psi_{1/2}^{hr} ( \mathbf r, \pi- \tilde \theta_{12}^r )
 \right \rbrace \Big \vert_{\varphi_2 \rightarrow \varphi_1, \, (z-L) \rightarrow z}\,.
\end{align} 
This amounts to flipping the signs of $\left \lbrace k_{z}^{(3/2),er }, \, k_{z}^{(1/2),er }, \,k_{z}^{(3/2),hr } ,\, k_{z}^{(1/2),hr } \right \rbrace $, which is because we need to consider here the left-moving electron-like and hole-like wavefunctions \cite{asano}. The ``left-moving'' wavefunctions are physically admissible in this region, because they are the ones which decay exponentially.

\end{enumerate}

Since the final results depend on the phase difference $\varphi_{12} = \varphi_2 -\varphi_1 $, for simplification of the notations, we can set $\varphi_1 = 0$ and $\varphi_2 = \varphi_{12} $, without any loss of generality.
Imposing the continuity of the wavefunction $\Psi$ at the junctions located at $z=0$ and $z=L$, we get the following conditions:
\begin{align}
\label{eqbdy}
\psi_{I}  (x, y, 0, k_\perp) = \psi_{II}  (x, y, 0, k_\perp) \text{ and } \psi_{II}  (x, y, L, k_\perp) 
= \psi_{III}  (x, y, L, k_\perp)\,.
\end{align}
From the eight components of the wavefunction, we get
$2\times 8=16 $ linear homogeneous equations in the $16$ variables $ \left( a_{32l}, \, a_{12l},\, b_{32l}, \,b_{12l}, \, a_{32}, \, a_{12},\, b_{32}, \,b_{12},\,
c_{32}, \, c_{12},\, d_{32}, \,d_{12}
, \, a_{32r}, \, a_{12r},\, b_{32r}, \,b_{12r} \right)$, which constitute the $16$ unknown coefficients of the piecewise-defined wavefunction. In these resulting equations, while the overall $z$-independent factors of $e^{ i \left( k_x\, x \, + \, k_y\, y \right)}$ totally cancel out, the phase factors introduced
by the action of $e^{- i \,\phi \, {\mathbb{1}_{2\times 2}} \otimes J_z}$ also cancel out component by component.
Let $M$ be the $16\times 16 $ matrix constructed from the coefficients of the 16 variables. The consistency of the equations is ensured by the condition $\text{det} M = 0 $. From this equation, we can determine the energy eigenvalues of the subgap ABSs, which are localized near the junctions with localization lengths $ \sim \kappa^{-1}_{1}$ and $ \sim \kappa^{-1}_{2}$ from the barrier, because they decay exponentially as we move away from the junction location into the superconducting region.

%%%%%%%%%%%%%%%%figure %%%%%%%%%%%%%%
\begin{figure}[t]
\centering
\subfigure[]{\includegraphics[width=0.4 \textwidth]{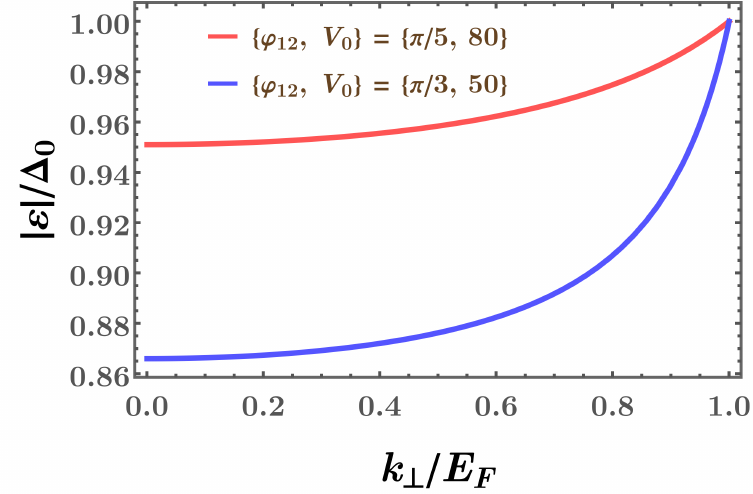}}\qquad
\subfigure[]{\includegraphics[width=0.4 \textwidth]{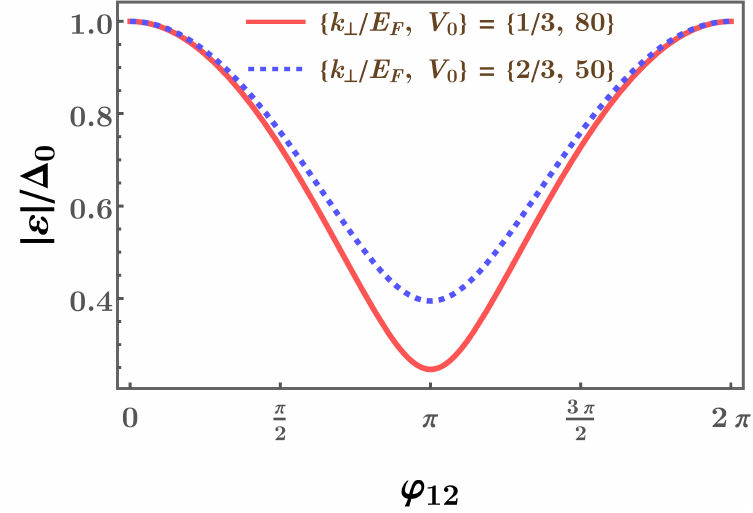}}
\caption{
Behaviour of $ |\varepsilon| $ against $k_\perp $ [subfigure (a)] and $\varphi_{12}$ [subfigure (b)] for graphene/Weyl-semimetal, using the expressions of Refs.~\cite{krish-moitri, debabrata-krish}. We have set $E_F=1$ and $L=0.01$ for all the curves, and the remaining parameters are shown in the plotlegends.
\label{fige_g}}
\end{figure}
%%%%%%%%%%%%%%%%%%%%%%%%%%%%%%%%%%%%%%%%%%%%%%%%%%%

%%%%%%%%%%%%%%%%%%%%%
\section{Results}
\label{secresults}

To simplify the calculations, instead of trying to compute the determinant of a $16 \times 16 $ matrix, we adopt the following strategy to obtain the solutions for $\varepsilon$ in the thin-barrier limit. Although this limit is equivalent to a Dirac delta potential, due to the fact that we do not have any constraint on the derivatives of the wavefunction across the junctions (due to the nature of the RSW Hamiltonian, which is linear in the position space
derivatives, when written in the position space representation), the
standard delta-function-potential approximation \cite{zagoskin,kwon_krish,ips-jj-sd} for thin barriers cannot be taken from the start \cite{krish-moitri}. Instead, we need to start with Eq.~\eqref{eqbdy}, and impose the appropriate limits in the expressions appearing in the equations obtained from the boundary conditions. Next, in region II, we employ the approximations
\begin{align}
k_{z}^{(3/2),e} \, L \rightarrow - \frac{2}{3 } \, \chi\,,\quad 
k_{z}^{(1/2),e} \, L \rightarrow - 2 \, \chi\,, \quad
k_{z}^{(3/2),h} \, L \rightarrow \frac{2}{3 } \, \chi\,,\text{ and }
k_{z}^{(1/2),h} \, L \rightarrow 2 \, \chi \,,
\end{align}
in the exponential factors representing plane waves propagating along the $z$-direction.
Furthermore, the $\epsilon $-dependence disappears from the angles, since $ -\theta_{32n} \simeq \tilde \theta_{32n} \simeq \arcsin \left(  \frac{V_0-E_F} {3\, k_\perp/ 2}\right) $ and $ -\theta_{12n} \simeq \tilde \theta_{12n} \simeq \arcsin \left(  \frac{V_0-E_F} { k_\perp/ 2}\right) $.

Plugging in the above approximations, we  solve for $\left(  a_{32}, \, a_{12},\, b_{32}, \,b_{12} \right) $ and $\left( c_{32}, \, c_{12},\, d_{32}, \,d_{12} \right) $, using the first four and the last four components of the matrix equation $\psi_{I}  (x, y, 0) = \psi_{II}  (x, y, 0)$, respectively, in terms of the remaining 8 variables. This is because, in region II, the last(first) four entries/components of each electron(hole) wavefunction are zero.
The resulting expressions are used to eliminate the normal state coefficients in the matrix equation $\psi_{II}  (x, y, L) 
= \psi_{III}  (x, y, L)$, and we end up with an $8 \times 8 $ matrix $\tilde M $ involving 8 independent variables. The values of $\varepsilon$ can now be obtained by demanding $\text{det} \tilde M = 0 $ for consistency. Even after adopting these simplifying steps, in the end, we have to deal with a quartic equation in the variable $  \eta_R \equiv  \exp(2\,i\,\beta ) $, with lengthy coefficients accompanying various powers of $  \eta_R $. Consequently, a simple analytic expression for $\varepsilon $ cannot be obtained, unlike semimetals having two bands \cite{krish-moitri,debabrata-krish}.

%%%%%%%%%%%%%%%%figure %%%%%%%%%%%%%%
\begin{figure}[t]
\centering
\subfigure[]{\includegraphics[width=0.35 \textwidth]{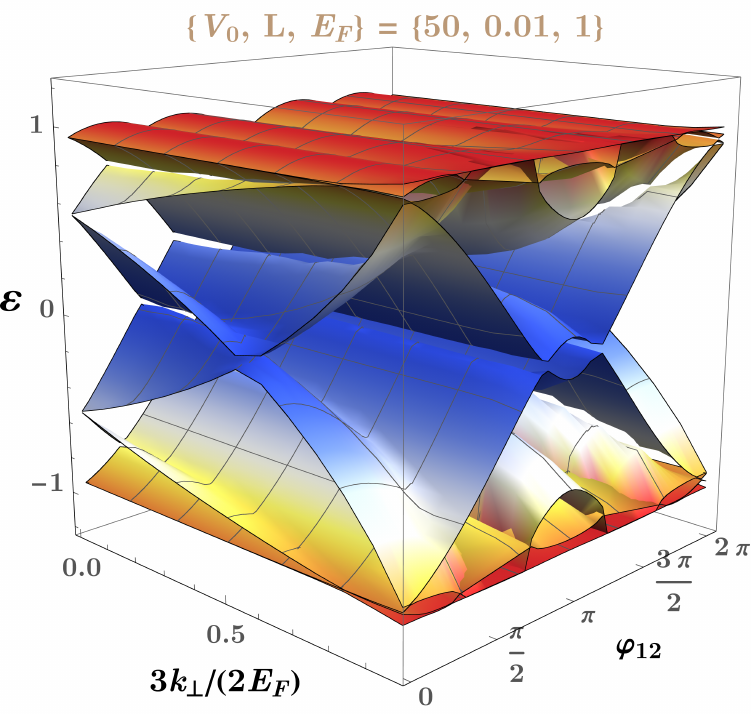}}\hspace{1 cm}
\subfigure[]{\includegraphics[width=0.475 \textwidth]{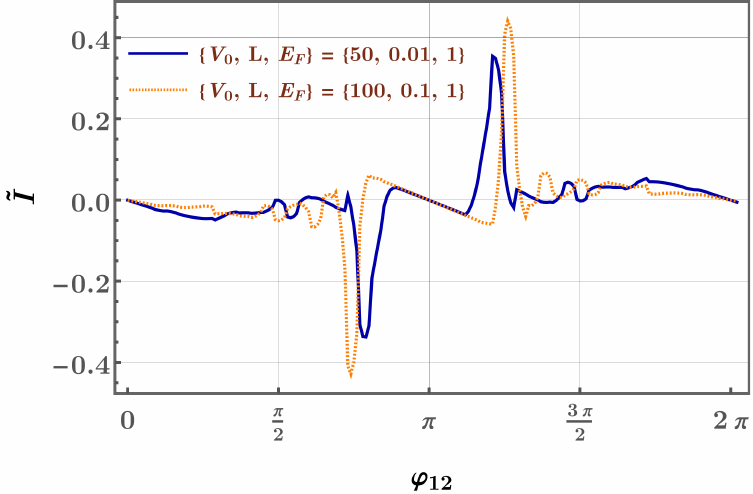}}
\caption{
(a) Energy $\varepsilon $ of the four pairs of Andreev bound states against the $k_\perp $-$\varphi_{12}$ plane, for $V_0 = 50$, $L=0.01$, and $E_F = 1$. (b) The behaviour of the total Josephson current ($\propto \tilde I $) in arbitrary units, as a function of $\varphi_{12}$, obtained at $k_B \, T = 0.005\, \Delta_0 $, for two sets of parameter values as shown in the plotlegends.
\label{figj}}
\end{figure}
%%%%%%%%%%%%%%%%%%%%%%%%%%%%%%%%%%%%%%%%%%%%%%%%%%%

As detailed above, we can determine the values of $\varepsilon $ only by finding the roots of the polynomials numerically, which are of quadratic order in $\sin(2 \beta )$ and
$\cos (2 \beta )$. Looking for real solutions from the real and imaginary components of the resulting complex-valued equation, we get upto four distinct values of $ |\varepsilon| $, for a given set of parameter values. This is because we have two real equations of quartic order in $\cos (2\beta )$ and $\sin (2\beta )$, whereas for each of the two-band models studied earlier, only a linear-order equation in $\cos (2\beta )$ had to be solved (which gave rise to only one pair of ABSs).
The energies of the subgap states appear as the pair $\pm |\varepsilon| $ for each value of $ |\varepsilon| $. In Fig.~\ref{fige}, we show their behaviour as functions of $ k_\perp $ (with a fixed value of $\varphi_{12}$) and $\varphi_{12}$ (with a fixed value of $  k_\perp $), for some representative values of $V_0$, $L$, and $E_F $. The bound state energies are periodic in $\varphi_{12}$ (with period $2\pi $) and are symmetric about the line $\varphi_{12} = \pi$. Fig.~\ref{figj}(a) illustrates the variation of the four pairs of $\varepsilon $-values against the $k_\perp$-$\varphi_{12} $ plane and, hence, shows the dependence of $\varepsilon $ on both these variables in a combined way.

In order to compare the bound-state energies with those found for two-band systems in Refs.~\cite{krish-moitri, debabrata-krish}, we use their explicit expressions (cf. Eq.~(20) of Ref.~\cite{krish-moitri} and Eq.~(13) of Ref.~\cite{debabrata-krish}) to plot $ |\varepsilon| $ against $k_\perp $ [subfigure (a)] and $\varphi_{12}$ [subfigure (b)] in Fig.~\ref{fige_g}. The values for the graphene and Weyl semimetal cases are identical, which is not surprising, because both harbour pseudospin-1/2 quasiparticles with isotropic linear-in-momentum dispersions.

The Josephson current density across the two junctions, at a temperature $T$, is given by \cite{zagoskin,titov-graphene}
\begin{align}
% debabrata paper
I_J(\varphi_{12}) = -\frac{2\, e}{\hbar}\,\frac{W^2} {(2\,\pi)^2} \sum_{n=1} ^ 8
\int  dk_x\,dk_y \, \frac{ \partial \epsilon_n}
{ \partial \varphi_{12}} \,f(\epsilon_n) \,,
\end{align}
where $\epsilon_n$ labels the energy values of the eight ABSs, and $f(\lambda ) = 1 / \left(  1 + e^{\frac{ \lambda }{k_B\, T}} \right)$ is the Fermi-Dirac distribution function. Fig.~\ref{figj}(b) shows the behaviour of $I_J$ as a function of $\varphi_{12}$, scaled by appropriate numbers/variables (this scaled quantity being denoted as $\tilde I $), for two sets of parameters.

%%%%%%%%%%%%%%%%%%%%
\section{Summary and outlook}
\label{secsum}

In this paper, we have considered an S-B-S sandwich configuration built with Rarita-Schwinger-Weyl semimetal, with the aim to determine the spectrum of ABSs in the thin-barrier limit. We have assumed a weak and homogeneous s-wave pairing in each superconducting region, which can be created via proximity effect \cite{proximity-sc} by placing a superconducting electrode near it. The barrier region can be implemented by applying a voltage of magnitude $V_0 $ across a piece of semimetal in its normal state. By using the appropriate BdG Hamiltonian, we have determined the wavefunction, which localizes at the boundaries, in a piecewise continuous manner. Enforcing consistency
of the equations obtained from matching the boundary conditions, we need to find the roots of the complex-valued polynomial in $e^{2\,i\,\beta}$, resulting from the vanishing of the relevant determinant. The solutions give the discrete energy spectrum $\varepsilon $ of the subgap Andreev states. Due to the higher order of the polynomial to be solved, one cannot find a closed-form analytical expression. Hence, we have solved for the admissible roots of the equation numerically, and have shown the results for some representative parameter values. As anticipated, in contrast with the two-band semimetals studied extensively so far, there exist multiple localized states (rather than two for two-band semimetals) in the thin-barrier limit. Furthermore, unlike the two-band semimetals, each value of $\varepsilon $ has a complicated dependence on the phase difference $\varphi_{12} $, which cannot be determined analytically. We have also derived the behaviour of the Josephson current, determined by the ABSs, and have illustrated it via a representative plot.

In future, it will be worthwhile to study a generalization of the isotropic version of the RSW semimetal studied here, where the full rotational symmetry of RSW is broken to the O$_h$ symmetry \cite{isobe-fu,ips-cd}, with the dispersion featuring anisotropic velocity parameters. An S-B-S junction set-up with such an anisotropic system is expected to show a richer structure of the ABSs, albeit with the need to solving more complicated equations. Another avenue to explore is to introduce a tilt in the band dispersion \cite{debabrata} and investigate the resulting ABSs.
Yet another interesting set-up is to consider scenarios where the dispersion is rotated about the $z$-axis across the junction(s), as considered in Refs.~\cite{tilted_jn,ips_jns}.
Lastly, RSW S-B-S junctions for higher angular momentum channels (e.g., d-wave symmetric pairing channel \cite{igor2}) and for FFLO pairings \cite{emil_jj_WSM,debabrata} are left for future investigations.

In our follow-up work~\cite{ips-jj-sd}, we have computed the energies of the ABSs emerging in 2d semi-Dirac semimetals (which feature a hybrid of linear and quadratic dispersions along the two mutually perpendicular momentum-axes), by considering the propagation of the quasiparticles/quasiholes along the quadratic-dispersion direction. Such hybrid dispersions  appear in the low-energy spectra of a tight-binding model on (1) the honeycomb lattice in a magnetic field (resulting in the
so-called Hofstadter spectrum), and (2) a square-lattice with three bands of spinless fermions.

%%%%%%%%%%%%%%%%%%%%%%%%%appendix%%%%%%

%%%%%%%%%%%%%%%%%
\bibliography{biblio_jj}
\end{document}